\documentclass[a4paper,12pt]{article}
\pdfoutput=1
\usepackage{a4wide}
\usepackage{amsmath,amsfonts,amssymb}
\usepackage{youngtab}
\usepackage{graphicx}
\usepackage{enumerate}
\usepackage{epsfig}
\usepackage{soul}
\usepackage{verbatim,setspace}
\usepackage{xparse}
\usepackage{tikz}
\usetikzlibrary{calc}
\usepackage{floatrow}

\DeclareDocumentCommand{\hcancel}{mO{0pt}O{0pt}O{0pt}O{0pt}}{%
    \tikz[baseline=(tocancel.base)]{
        \node[inner sep=0pt,outer sep=0pt] (tocancel) {#1};
        \draw[black] ($(tocancel.south west)+(#2,#3)$) -- ($(tocancel.north east)+(#4,#5)$);
    }
}

\newcommand{\cL}{{\cal L}}

\newcommand{\ov}{\overline}

\newcommand{\be}{\begin{equation}}
\newcommand{\ee}{\end{equation}}
\newcommand{\bea}{\begin{eqnarray}}
\newcommand{\eea}{\end{eqnarray}}

\DeclareMathSymbol{\mg}{\mathrel}{symbols}{"1D}

\newcommand{\MET}{E\llap{/\kern1.5pt}_T}
\newcommand{\nn}{\nonumber}

\def\nn{\nonumber}
\def\bra{\langle}
\def\ket{\rangle}

\def\beq{\begin{equation}}
\def\eeq{\end{equation}}

\def\psqr#1#2{{\vcenter{\vbox{\hrule height.#2pt
        \hbox{\vrule width.#2pt height#1pt \kern#1pt
        \vrule width.#2pt}
        \hrule height.#2pt \hrule height.#2pt
        \hbox{\vrule width.#2pt height#1pt \kern#1pt
        \vrule width.#2pt}
        \hrule height.#2pt}}}}
\def\sqr#1#2{{\vcenter{\vbox{\hrule height.#2pt
        \hbox{\vrule width.#2pt height#1pt \kern#1pt
        \vrule width.#2pt}
        \hrule height.#2pt}}}}

\newcommand{\C}[1]{\mathcal{#1}}

\def\ov{\overline}

\newcounter{oldcounter}
\addtocounter{equation}{1}
\begin{document}

\thispagestyle{empty}

\begin{center}
\begin{spacing}{2.5}
{\Huge{\bf Singlet-Doublet Model:}}

{\Huge{\bf Dark matter searches}}

{\Huge{\bf  and LHC constraints }}

\end{spacing}

{\bf\large Lorenzo Calibbi$^{a,b}$, Alberto Mariotti$^{c}$, Pantelis Tziveloglou$^{c}$}


\vspace{0.3cm}
\noindent \textit{{\small 
$^a$State Key Laboratory of Theoretical Physics, Institute of Theoretical Physics,\\Chinese Academy of Sciences, 
Beijing 100190, P.~R.~China\\
$^b$Service de Physique Th\'eorique, Universit\'e Libre de Bruxelles,\\C.P. 225, B-1050, Brussels, Belgium\\
$^c$Theoretische Natuurkunde and IIHE/ELEM, Vrije Universiteit Brussel,\\
and International Solvay Institutes, Pleinlaan 2, B-1050 Brussels, Belgium \\
}}

\end{center}
\vspace{1cm}
 \begin{abstract}
\noindent \normalsize
The Singlet-Doublet model of dark matter is a minimal extension of the Standard Model with dark matter that is a mixture of a singlet and a non-chiral pair of electroweak doublet fermions.  The stability of dark matter is ensured by the typical parity symmetry, and, similar to a `Bino-Higgsino' system, the extra matter content improves gauge coupling unification.
We revisit the experimental constraints on the Singlet-Doublet dark matter model, combining the most relevant bounds from direct (spin independent and spin dependent) and indirect searches.
We show that such comprehensive analysis sets strong constraints on a large part of the 4-dimensional parameter space, 
closing the notorious `blind-spots' of spin independent direct searches. 
Our results emphasise the complementarity of direct and indirect searches in probing dark matter models in diverse mass scale regimes. We also discuss the LHC bounds on such scenario, which play a relevant role in the low mass region of the dark matter candidate.

\end{abstract}

\newpage

\tableofcontents

\setcounter{page}{1}

\section{Introduction}

In defiance of a plethora of experimental searches, the mass and dynamics of Dark Matter (DM) remain largely unknown. Up to now, there is no indication that the DM particle interacts with the known matter other than gravitationally. However, if one assumes that the relic abundance of DM today is determined by the same physical processes that determined the relic abundance of baryonic matter, one would require that DM at some early point in the history of the universe was in thermal equilibrium with the Standard Model (SM) particles.

The idea of a thermal relic DM had been further motivated by the following observation: A particle whose mass and dynamics are dictated by electroweak (EW) physics would have a thermal evolution that can reproduce the observed DM relic density today. In simple constructions of this type, the DM has spin independent interactions with the nucleons via a $Z$-boson exchange that are orders of magnitude larger than experimental limits. Another popular proposal that relates the mass and dynamics of DM with the electroweak sector is the neutralino of supersymmetric extensions of the SM. There, the mass of the DM candidate is related to the supersymmetry breaking scale, which in turn is related to the EW scale by means of the naturalness criterion. In this case the DM is a Majorana particle so that the spin independent DM-nucleon interaction via $Z$-boson exchange is absent.

The negative searches in DM and collider experiments for these explicit scenaria has led to the consideration of different approaches in describing the properties of DM. One direction that has been extensively studied is by using the tools of effective field theory, where the coupling of DM with SM particles is parametrised by higher dimensional operators suppressed by the scale that determines the validity of the effective theory. While this approach has been very useful in obtaining model independent bounds on DM properties from direct and indirect searches, it has been shown \cite{Shoemaker:2011vi,Buchmueller:2013dya,Busoni:2013lha} that it can be misleading when considering the complementarity of these searches with DM searches in high-energy colliders.

In view of this difficulty, simplified Models of DM have been lately motivated as an alternative to effective DM theories that allow for a unified study of the multiple DM experimental fronts \cite{Shoemaker:2011vi,Malik:2014ggr,Abdallah:2014hon}. Furthermore, the negative LHC searches for complete ultraviolet extensions of the SM with a natural DM candidate (e.g. supersymmetry), point towards model building beyond the schemes that specific UV completions impose. 
In this perspective, a well motivated simplified Model of DM can capture the dynamics of different UV completions  that are relevant for each specific experimental search.

In this paper we discuss the theory and phenomenology of a Simplified Model of DM that consists of a singlet and an electroweak doublet fermion \cite{Mahbubani:2005pt,Cohen:2011ec,Cheung:2013dua},
besides the SM field content. The dark matter candidate of the `Singlet-Doublet' Model (SDM) is identified as the lightest neutral fermion of the new sector whose stability is ensured by a ${\bf{Z}_2}$ symmetry.

Within the vast landscape of DM models, the SDM enjoys certain motivating features: i) Given the tight experimental constraints on chiral dark matter, it represents a minimal model of electroweak dark matter that is stable by a symmetry of the theory.\footnote{For a similar construction with an electroweak doublet and a triplet, see \cite{Dedes:2014hga}.} ii) The sensitivity of current direct search experiments is high enough that DM-nucleon interactions via a Higgs boson exchange are now being probed. The SDM is a simplified model that can be used to parametrise these bounds since the DM candidate has tree-level interactions with the Higgs boson. iii) The new field content of this model improves the unification of gauge couplings with respect to the Standard Model (however the unification is reached at a scale which is too low to avoid proton decay. A way to address this issue has been proposed in \cite{Mahbubani:2005pt}). iv) The SDM captures the relevant dynamics of well-studied supersymmetric candidates such as the Bino-Higgsino system in MSSM and the Singlino-Higgsino system in NMSSM-type scenarios.

Many aspects of the theory and phenomenology of SDM have been investigated in the past \cite{Cohen:2011ec,Cynolter:2008ea,Eramo:2007ab,Enberg:2007rp,Cohen:2010gj,Joglekar:2012vc,Abe:2014gua}, including the recent analysis of \cite{Cheung:2013dua} where the constraints on the parameter space have been updated focusing on spin independent direct detection experiments.

The goal of the present work is to explore the complementarity of dark matter and collider searches for the SDM, especially in view of the vast progress that has been recently achieved in the indirect search observational front. In particular, we see how a combination of spin dependent and indirect searches probes regions of the parameter space where spin independent searches lose sensitivity, the so called `blind spots'. Combining these bounds, one can reach sensitivities that allow for exclusion of a very large part of the parameter space, extending up to DM mass values of the order TeV.

We also focus on the special features that the model exhibits when the DM candidate is relatively light, in particular on the extra constraints coming from EW precision observables as well as $Z$-boson and Higgs boson decay width. In particular the LHC constraint on Higgs invisible decays has a relevant impact on previously unbounded regions of the parameter space. We will observe that the low mass scenarios with relatively large Higgs-DM interaction are almost completely ruled out, except for small corners of the parameter space that will be accessible at the next LHC run.

The structure of the paper is the following: In section \ref{sec:model} we introduce the model, we discuss the relevant DM couplings to SM particles and we explain the main DM annihilation channels. In section \ref{sec:results} we display the limits set on the parameter space of the model by indirect and direct DM searches, showing their complementarity. In section \ref{sec:lhc} we focus on the low mass region and show the interplay of DM searches and collider constraints and finally, in section \ref{summary}, we summarise our results.

\section{The Model}
\label{sec:model}

We now introduce the Singlet-Doublet Dark Matter model (SDM) that we will study in the rest of the paper.
The field content of the dark sector consists of new Weyl fermions with the following transformation properties under U(1) and SU(2):
\be
\label{eq:lagrangian}
\lambda:\, ({\bf 1},0)\,,\quad \psi_1=\left(\begin{array}{c}\!\!\!\psi_1^0 \!\!\!\\\!\!\!\psi_1^-\!\!\!\end{array}\right):\, ({\bf 2},-1)\,,\quad \psi_2=\left(\begin{array}{c}\!\!\!\psi_2^+ \!\!\!\\\!\!\!\psi_2^0\!\!\!\end{array}\right):\, ({\bf 2},1)
\ee
The electroweak doublets have opposite hypercharge in order to cancel gauge anomalies as well as to avoid the strong bounds on chiral dark matter \cite{Goodman:1984dc}. Furthermore, the new fields are assumed to be odd under a ${\bf Z}_{2}$ symmetry under which all SM fields are even. With this field content, apart from the obvious couplings to gauge bosons from the covariant derivative, interaction with the SM via renormalisable operators is achieved exclusively by coupling to the Higgs. The dynamics of the new fields are then governed by the following Lagrangian:
\bea\label{Lf}
\cL_{SDM}&=& {i\over 2}\left(\,\ov{\lambda}\sigma^\mu\partial_\mu\lambda + \,\ov{\psi}_1 \sigma^\mu D_\mu\psi_1+ \,\ov{\psi}_2 \sigma^\mu D_\mu\psi_2\right) -{m_{\textrm{\tiny S}}\over 2}\lambda\lambda-m_{\textrm{\tiny D}}\,\psi_1\cdot\psi_2\nn
\\
&&-y_1\,\psi_1\!\cdot\!H\, \lambda-y_2\,\ov{\psi}_2\,\ov{\lambda}\,H+h.c.\,,
\eea
and determined by the four parameters
\be
m_{\textrm{\tiny S}},~m_{\textrm{\tiny D}},~y_1,~y_2,
\label{parameters}
\ee
denoting respectively the mass terms of the singlet $\lambda$, the doublets $\psi_{1,2}$ and the Yukawa couplings among these fermions and the SM Higgs doublet $H$ which can also be expressed as
\be
y_1=y\cos\theta,\quad y_2=y\sin\theta\,.
\ee
In our analysis, we will neglect the overall CP violating phase of these parameters and take them real, while, without loss of generality we also take $m_{\textrm{\tiny S}},\,m_{\textrm{\tiny D}},\,y>0$. For constraints on SDM from limits on the electron EDM $d_e$ in the case of a non-vanishing CP violating phase, see \cite{Mahbubani:2005pt}.\footnote{Considering the recent updated limit $|d_e|<8.7\times10^{-29}~e{\rm cm}$ \cite{Baron:2013eja}, 
we do not
expect the improved limit on the single physical phase (that can be chosen to be that of $m_{\textrm{\tiny D}}$) to be below $\mathcal{O}(0.1)$ 
even for $m_{\textrm{\tiny S}},\,m_{\textrm{\tiny D}}$ of $\mathcal{O}(100)$ GeV.}

Upon electroweak symmetry breaking, the neutral components of the doublets and the singlet mix. The physical spectrum of the model then consists of three Majorana fermions with increasing mass $m_{1,2,3}$ and a Dirac charged fermion whose mass we denote as $m_{\psi^+}$. At tree level, one has  $m_{\psi^+}= -m_{\textrm{\tiny D}}$. The mass matrix of the neutral states and the rotation to the mass eigenstates are given by
\be
\label{eq:mass_matrix}
\mathbf{M}=\left(\begin{array}{ccc}m_{\textrm{\tiny S}} & {y_1v\over \sqrt{2}} & {y_2v\over \sqrt{2}} \\{y_1v\over \sqrt{2}} & 0 & m_{\textrm{\tiny D}} \\{y_2v\over \sqrt{2}}&m_{\textrm{\tiny D}} & 0\end{array}\right)\,,\quad
\left(\!\!\!\begin{array}{c}\chi_1 \\\chi_2 \\\chi_3\end{array}\!\!\!\right)=\mathbf{U}\left(\!\!\!\begin{array}{c}\lambda \\\psi_1^0 \\\psi_2^0\end{array}\!\!\!\right)\,.
\ee
The lightest eigenstate $\chi_1$ is our dark matter candidate.\footnote{We use $\chi$ and $\chi_1$ (as well as $m_1$ and $m_\chi$) interchangeably.}
Its composition in terms of the fields in the gauge basis is 
\be
\chi_1=U_{11}\lambda+U_{12}\psi_1^0+U_{13}\psi_2^0, \quad |U_{11}|^{2}+|U_{12}|^{2}+|U_{13}|^{2}=1.
\ee

The dark sector interacts with the SM via couplings to the W, the Z and the Higgs bosons. Written in the mass eigenstate basis and standard 4-component notation,\footnote{We have $X_n=\left(\chi_{n\alpha}~\ov{\chi}_n^{\dot{\alpha}}\right)^T$ and 
$\Psi^+=\left(\psi_{2\alpha}^+~\ov{\psi^-_{1}}^{\dot{\alpha}}\right)^T$.}
they are given by
\bea
\cL&\supset& -h\ov{X}_n(c_{h\chi_m\chi_n}P_L+c_{h\chi_m\chi_n}^*P_R)X_m-Z_\mu\ov{X}_m\gamma^\mu(c_{Z\chi_m\chi_n}P_L-c_{Z\chi_m\chi_n}^*P_R)X_n\nn
\\
&&-{g\over \sqrt{2}}(U_{n3}W_\mu^-\ov{X}_n\gamma^\mu P_L\Psi^+ -U_{n2}^*W_\mu^-\ov{X}_n\gamma^\mu P_R\Psi^++h.c.)\,,
\label{eq:couplings}
\eea
where
\be\label{zetaeta}
c_{Z\chi_m\chi_n}\!=\!{g\over 4c_W}(U_{m3}U_{n3}^*-U_{m2}U_{n2}^*)\,,\ c_{h\chi_m\chi_n}\!=\!{1\over \sqrt{2}}(y_1U_{m2}^*U_{n1}^*+y_2U_{m3}^*U_{n1}^*)\,.
\ee
The couplings of the DM candidate then take the simple form
\bea\label{Lchi1}
\cL_{\chi_1}&\supset& -{c_{h\chi\chi}\over 2} h\ov{X}_1X_1-c_{Z\chi\chi} Z_\mu\ov{X}_1\gamma^\mu\gamma^5X_1\nn
\\
&&-{g\over \sqrt{2}}(U_{13}W_\mu^-\ov{X}_1\gamma^\mu P_L\Psi^+ -U_{12}^*W_\mu^-\ov{X}_1\gamma^\mu P_R\Psi^++h.c.)\,,
\eea
where $c_{h\chi\chi}$ and $c_{Z\chi\chi}$ can be written as
\bea
\label{eq:dm_couplings1}
&&c_{h\chi\chi}=-{(2y_1y_2m_{\textrm{\tiny D}}+(y_1^2+y_2^2)m_1)v\over m_{\textrm{\tiny D}}^2+(y_1^2+y_2^2){v^2\over 2}+2m_{\textrm{\tiny S}}\,m_1-3m_1^{2}}\,,
\\
&&c_{Z\chi\chi}=-{m_Zv(y_1^2-y_2^2)(m_1^{2}-m_{\textrm{\tiny D}}^2) \over 2(m_1^{2}-m_{\textrm{\tiny D}}^2)^2+v^2\left( 4y_1\,y_2\,m_1\,m_{\textrm{\tiny D}}+(y_1^2+y_2^2)(m_1^{2}+m_{\textrm{\tiny D}}^2) \right)}\,.
\label{eq:dm_couplings2}
\eea
We note that $m_1$ here can have either sign. From these expressions we observe that there are corners of the parameter space where the above couplings vanish. Such ``blind spots" of the Higgs \cite{Cheung:2013dua,Cohen:2011ec} and $Z$ boson couplings to dark matter occur for
\bea
\sin(2\theta)m_{\textrm{\tiny D}}+m_1=0\,\quad  &\Longrightarrow & \quad  c_{h\chi\chi}=0\,; \\
\quad |y_1|=|y_2|~~\mathrm{or}~~|m_1|=m_{\textrm{\tiny D}}\quad &\Longrightarrow & \quad   c_{Z\chi\chi}=0\,.
\label{blind}
\eea
Regarding the coupling to the $Z$ boson, the blind spot at $|y_1|=|y_2|$ is easily understood by looking at $c_{Z\chi\chi}$ of eq.~(\ref{eq:dm_couplings2}): when the two Yukawa couplings are equal, $|U_{12}|^2=|U_{13}|^2$ so that the coupling vanishes. This is analogous to the vanishing of the neutralino coupling to the $Z$
at $\tan\beta=1$ in supersymmetric scenarios. The second possibility, $|m_1|=m_{\textrm{\tiny D}}$, corresponds to the limit where
DM is almost purely made of the neutral components of the doublets, thus again to a case where the two mixing angles are almost equal 
$|U_{12}|^2\simeq |U_{13}|^2\simeq1/2$. In other words, in this second case $c_{Z\chi\chi}$ vanishes as DM becomes a Dirac fermion and the vector interaction of standard Dirac DM is restored.

Finally, we remind that the Lagrangian of eq.~(\ref{eq:lagrangian}) 
represents a generalisation of the Bino-Higgsino sector of the MSSM in the decoupling limit of only one light SM-like Higgs boson.
In the MSSM, supersymmetry dictates that the Yukawa couplings $y_1$ and $y_2$ are related to the U(1) gauge coupling $g'$. 
Instead, we take them here as free parameters. The MSSM limit of the model can be obtained as follows:
\be
y \rightarrow \frac{g'}{\sqrt{2}},\qquad  m_{\textrm{\tiny S}} \rightarrow M_1, \qquad  m_{\textrm{\tiny D}} \rightarrow - \mu, \qquad
 \cos \theta \rightarrow -\cos \beta, \qquad \sin \theta  \rightarrow \sin \beta,\nn
\ee
where $M_1$ is the SUSY-breaking Bino mass term, $\mu$ is the Higgsino mass term and $\tan\beta$ the ratio of the vevs
of the two Higgs doublets.

\subsection{Dark Matter annihilation channels}
\label{sec:annihilation}

Before we move on with the discussion of the constraints to SDM, we give here a brief overview of the main annihilation modes of the DM candidate. 
The annihilation modes participate both in the determination of the thermal relic density in the early universe and in the 
expected indirect detection rate in the late universe. 

In the following, for each annihilation channel, we focus in particular on the corresponding contribution for indirect detection.
We note that the annihilation of $\chi_1$ to fermions in the late universe can only be mediated by a $Z$ boson, since the s-wave amplitude of the Higgs-mediated contribution vanishes. Due to the helicity structure, the s-wave contribution to the $Z$-mediated annihilation cross section scales with the fermion mass. Therefore, in summary, the main annihilation channels in the SDM are to bosons and for high enough DM mass, to $t\ov{t}$.

\subsubsection*{$\chi\chi\to W^+W^-,\, ZZ$}
%
Dark matter annihilation into $W^+W^-$ occurs at tree level via a t-channel exchange of the charged fermion $\psi^{\pm}$ and via an s-channel exchange of a $Z$ or a Higgs boson. However, the former dominates the rate relevant for indirect detection since the $Z$/$h$ exchange has no s-wave contribution. Annihilation into $ZZ$ occurs at tree level via a t-channel exchange of a neutral fermion $\chi_{i}$ and via an s-channel exchange of a Higgs boson. As before, only the t-channel is relevant for indirect detection since the s-channel does not contribute in the vanishing velocity limit. Sommerfeld enhancement brings corrections to the tree level rate in the limit $m_{\chi}\gg m_W$. From the results of \cite{Hisano:2004ds}, we expect that the enhancement appears for $\chi_1$ that is almost purely doublet and is marginal in the parameter space we consider. We will therefore for simplicity neglect this effect.
%
\subsubsection*{$\chi\chi\to t\ov{t}$}
%
Dark matter annihilation to fermions occurs in an s-channel, through a $Z$ or a Higgs boson. Because of the helicity structure of the process, the Higgs channel is irrelevant for indirect searches while the s-wave amplitude for the $Z$ channel scales with the fermion mass so that, for heavy enough dark matter, the annihilation to two top-quarks is the dominant one.
%
\subsubsection*{$\chi\chi\to Z h$}
%
This process proceeds via t-channel exchange of a neutral fermion $\chi_i$ and via s-channel through a $Z$ boson. Both channels can potentially be relevant for indirect searches since none of them is velocity suppressed. Since the t-channel passes through a $\chi_i\chi_j h$ vertex, we expect that the rate of this process grows with $y$ (unless it falls on a blind spot).
%
\subsubsection*{$\chi\chi\to \gamma\gamma,\, \gamma Z$}
%
Annihilation into two photons or a photon and a $Z$ boson occurs at one loop via box diagrams of $W$ bosons and $\psi^{\pm}$ and via an s-channel  $Z$ and a triangle loop of SM fermions, $\psi^{\pm}$ or $W$ bosons. The calculation of the loop induced annihilation rate of DM to $\gamma\gamma$ or $\gamma Z$ in the MSSM context with a neutralino DM candidate was done in \cite{Bern:1997ng,Bergstrom:1997fh,Ullio:1997ke}. It was shown that for Higgsino DM heavier than around 300 GeV, the cross section for both channels reaches a plateau at $\sim10^{-28}\,$cm$^3$s$^{-1}$ while the non-perturbative, non-relativistic corrections introduce a resonant peak at $m_\chi\sim 6\,$TeV \cite{Boudjema:2005hb}.\footnote{In the parameter space that we consider in our numerical analysis in Section \ref{sec:results} (with $m_{\textrm{\tiny D}},~m_{\textrm{\tiny S}} < 5$~TeV), the non-perturbative enhancement can be at most of one order of magnitude in the annihilation cross section (see Fig. 5 of \cite{Boudjema:2005hb}).}

Other annihilation channels include $\chi\chi\to hh$ and $\chi\chi\to f\ov{f}$, where $f$ are SM fermions other than the top. Since the DM particles are Majorana fermions, the first one is again velocity suppressed and thus insignificant for indirect searches. As mentioned above, the annihilation to fermions pairs, in particular $b\ov{b}$ and $\tau^+\tau^-$, is mediated through a $Z$ boson and helicity suppressed, therefore possibly relevant only for DM masses below $m_W$. 
 
\section{Limits from Direct and Indirect Searches}
\label{sec:results}

In this section, we use a collection of direct and indirect searches for dark matter to put constraints on the parameter space of SDM. The bounds from spin-independent direct detection have been recently investigated in \cite{Cheung:2013dua} where it was shown that, even if they are strong for relatively large Yukawa couplings, they are ineffective around the `blind spot' regions, where the coupling of $\chi_1$ to Higgs vanishes. We will see that, after taking into account a combination of spin-dependent and indirect searches, these blind spot regions can be probed and excluded.

Throughout our analysis, the thermally averaged annihilation cross section of $\chi_1$ into the channels discussed in the previous section have been computed 
by modifying appropriately the expressions that can be found 
for example in \cite{Jungman:1995df}. We cross-checked our analytical calculations employing micrOMEGAs \cite{Belanger:2013oya}, finding an excellent agreement.

In the following analysis, we assume that the lightest state $\chi_1$ accounts for the whole dark matter abundance. Nevertheless, we show in the plots a line where its relic density $\Omega_\chi$ as computed by micrOMEGAs agrees with the Planck observations, $\Omega_{\rm DM}h^2\simeq 0.12$ \cite{Ade:2013zuv}. Such a line divides the parameter space in two regions, where $\chi_1$ as standard thermal relic would be respectively under and overabundant. Thus, $\chi_1$ can be considered a good DM candidate (and the bounds we are going to show apply) in the former case if a non-thermal production mechanism is at work in the early universe, while in the latter case a non-standard thermal history of the universe, providing DM dilution, has to be assumed. If, on the contrary, one sticks on $\chi_1$ as a standard thermal relic, the areas of our plots leading to overabundance would be excluded by the results of Planck on $\Omega_{DM} h^2$, while in those with underabundance $\chi_1$ would be interpreted as a subdominant component of DM and the sensitivity of DM experiments to our model would decrease due to the reduced $\chi_1$ density. We will comment about this possibility when discussing our results.

\subsection{Experimental input}

In our analysis, we use recent results by the Fermi-LAT \cite{Ackermann:2015zua}, PICO \cite{Amole:2015lsj}, IceCube \cite{Aartsen:2012kia} and LUX \cite{Akerib:2013tjd} collaborations, including prospects for the CTA experiment \cite{Silverwood:2014yza}. Before we move on with the results, we briefly report a few details about these searches, starting from searches for gamma rays.

The gamma ray flux obtained from annihilation of DM is given by
\be
\Phi_{\gamma}(\Delta\Omega)={1\over 8\pi}{\bra \sigma_A v\ket\over m_\chi^2}\int_{E_{min}}^{E_{max}}\sum_iB_i{dN^i_{\gamma}\over dE_{\gamma}}dE_{\gamma}\times\, \underbrace{\int_{l.o.s.}\int_{\Delta\Omega}\rho_{DM}^2(l,\Omega)\,dl\,d\Omega}_\text{\normalsize J}\,,
\ee
where ${\bra \sigma_A v\ket}$ is the velocity averaged total DM annihilation cross section, $m_\chi$ is the DM mass and $B_idN^i_\gamma/dE_\gamma$ is the differential gamma ray yield for a particular final state multiplied by the corresponding branching fraction. The J-factor encodes the information about the astrophysical density distribution of DM at the source. It is defined as the integral along the light of sight (l.o.s.) of the square of the density of the DM distribution in the source, integrated over the solid angle of observation. Under the assumption of a particular DM density distribution profile, the flux upper limits can be used to set upper bounds on $\bra \sigma_A v\ket$. Dwarf spheroidal galaxies have the advantage that the integrated J-factor is relatively insensitive to the choice of a DM density profile \cite{Strigari:2007at,Ackermann:2013yva}. The bounds on the annihilation cross sections of $\chi$ are obtained from the recent results of Fermi-LAT \cite{Ackermann:2015zua} where the negative search of gamma rays coming from 15 dwarf spheroidal satellite galaxies of the Milky Way is used to set upper limits on the DM annihilation cross section into certain channels, namely $e^+e^-$, $\mu^+\mu^-$, $\tau^+\tau^-$, $u\ov{u}$, $b\ov{b}$ and $WW$.

Searches for antiprotons have shown to set limits on $\chi\chi\to WW/ZZ$ (which are in most of the parameter space the dominant annihilation channels) that are either smaller or comparable to the ones set from $\gamma$-rays, depending on astrophysical details of their propagation through the galaxy \cite{Belanger:2012ta,Giesen:2015ufa,Jin:2015sqa}. We do not consider these searches in the present study, however, it is certainly interesting to explore to what extend these searches can improve the bounds presented here.

Concerning direct DM searches, we employ bounds on the spin-independent (SI) $\chi$-nucleon scattering cross section $\sigma^{SI}_{N\chi}$ and the spin-dependent (SD) $\chi$-proton scattering cross section $\sigma^{SD}_{p\chi}$ coming from LUX \cite{Akerib:2013tjd}, PICO \cite{Amole:2015lsj} and IceCube \cite{Aartsen:2012kia}. The latter one is an indirect, model-dependent limit on $\sigma^{SD}_{p\chi}$ obtained by interpreting the bounds on the flux of neutrinos from DM annihilation in the sun. Two different upper limits are given in \cite{Aartsen:2012kia}, a stronger one by assuming DM annihilation to $WW$ (resulting to neutrinos with a harder spectrum) and a weaker one by assuming annihilation to $b\ov{b}$ (resulting to neutrinos with a softer spectrum). In our case, for $m_\chi>m_W$, apart from certain regions of the parameter space where DM is mostly singlet, the dominant annihilation leads to hard neutrino spectra. One can recast the IceCube bounds for multiple annihilation channels with less or more hard neutrino spectra. Instead, in this work we take a conservative approach and apply these bounds only when the annihilation 
to hard spectra is dominant  \cite{Wikstrom:2009kw}, i.e.: $\langle\sigma v\rangle_{WW}+\langle\sigma v\rangle_{ZZ}+\langle\sigma v\rangle_{t\ov{t}}>10\,\langle\sigma v\rangle_{b\ov{b}}$.

\subsection{Results}
\label{results}

\begin{figure}[t!]
\begin{center}
\floatbox[{\capbeside\thisfloatsetup{capbesideposition={right,top},capbesidewidth=6.3cm}}]{figure}[\FBwidth]
{\caption{Limits and CTA prospects on SDM for $y=0.01$ and any $\tan\theta$. The red region is excluded by 
Fermi-LAT limit \cite{Ackermann:2015zua} on $\langle\sigma v\rangle_{WW}$.
The prospective exclusion by CTA (from \cite{Silverwood:2014yza}) is denoted by the dashed grey line. The excluded regions are drawn under the assumption that $\Omega_{\chi}$ matches the observed value (see text for details).
Regions where $\chi$ would be overabundant (underabundant) are denoted by ``over" (``under").}}
{\includegraphics[width=8cm]{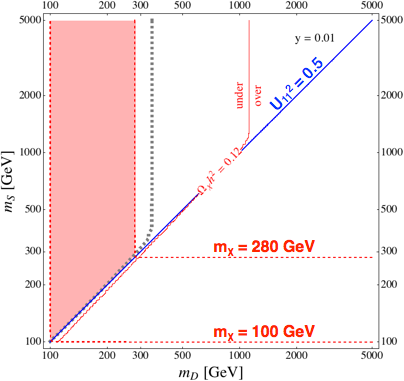}\label{SDfer_y001}}
\end{center}
\end{figure}

\begin{figure}[t!]
\includegraphics[width=7cm]{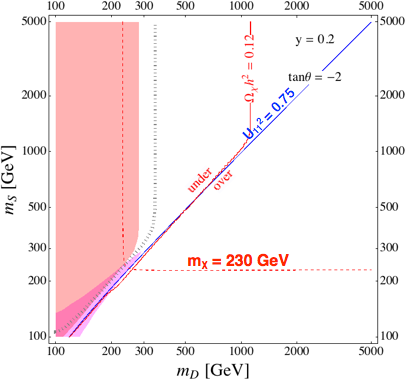}
\hspace{0.5cm}
\includegraphics[width=7cm]{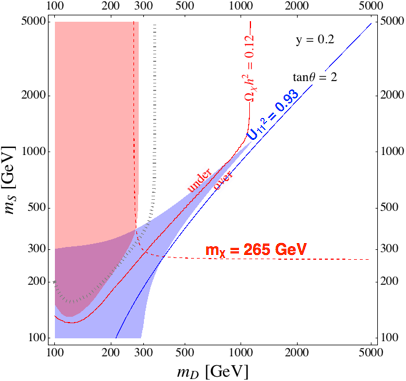}
\includegraphics[width=7cm]{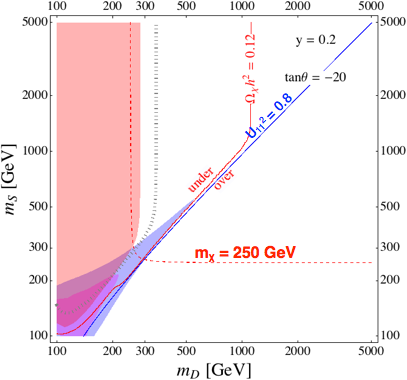}
\hspace{0.5cm}
\includegraphics[width=7cm]{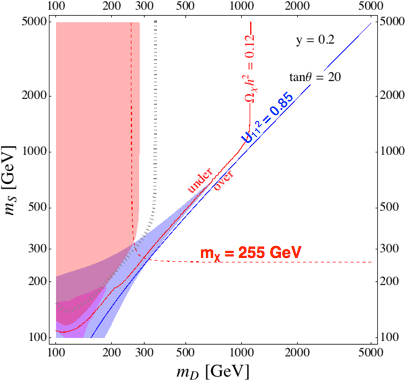}
\caption{Limits and CTA prospects on SDM for $y=0.2$ and various values of $\tan\theta$. Red: region excluded by $\langle\sigma v\rangle_{WW}$ \cite{Ackermann:2015zua}. Blue: exclusion from direct detection limits on $\sigma_{N\chi}^{SI}$ \cite{Akerib:2013tjd}. 
Magenta: exclusion from IceCube limits on $\sigma_{p\chi}^{SD}$ \cite{Aartsen:2012kia}.
CTA sensitivity prospect is shown as a dashed grey line. The excluded regions are drawn under the assumption that $\Omega_{\chi}$ matches the observed value (see text for details).
Regions where $\chi$ would be overabundant (underabundant) are denoted by ``over" (``under").}
\label{SDfer_y02}
\end{figure}

\begin{figure}[t!]
\includegraphics[width=7cm]{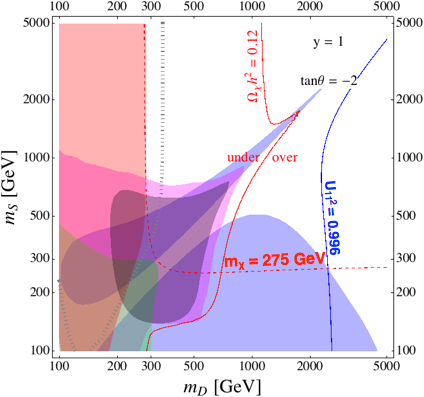}
\hspace{0.5cm}
\includegraphics[width=7cm]{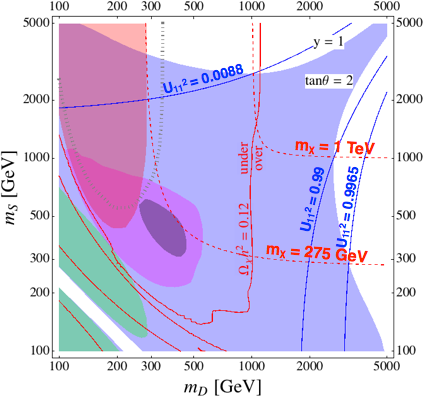}
\includegraphics[width=7cm]{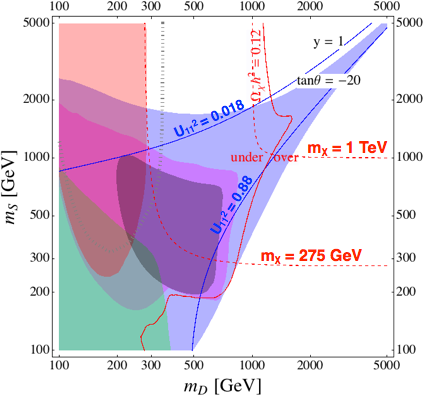}
\hspace{0.5cm}
\includegraphics[width=7cm]{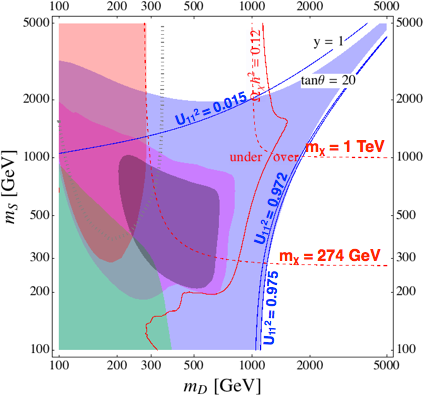}
\caption{The same as before for $y=1$. 
Additional excluded regions that appear are the following.
Green: exclusion from direct detection limits on $\sigma_{p\chi}^{SD}$  \cite{Amole:2015lsj}. Purple: region excluded by
$\langle\sigma v\rangle_{t\ov{t}}$ \cite{Ackermann:2015zua}.}
\label{SDfer_y1}
\end{figure}

\begin{figure}[t!]
\includegraphics[width=7cm]{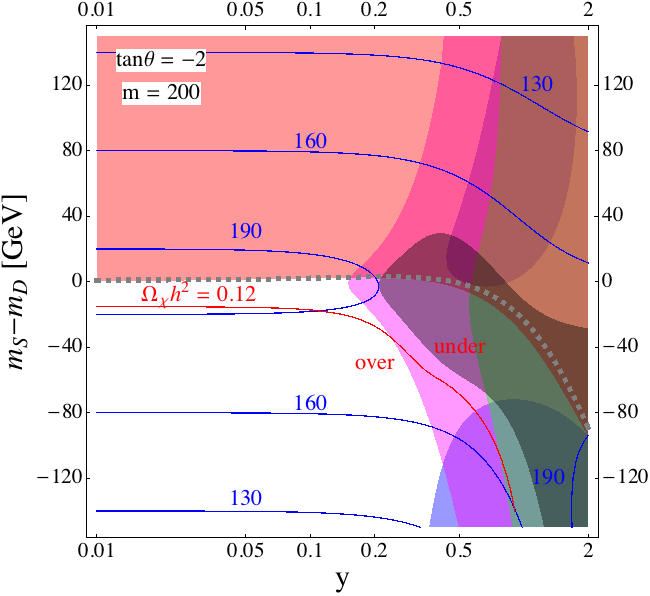}
\hspace{0.5cm}
\includegraphics[width=7cm]{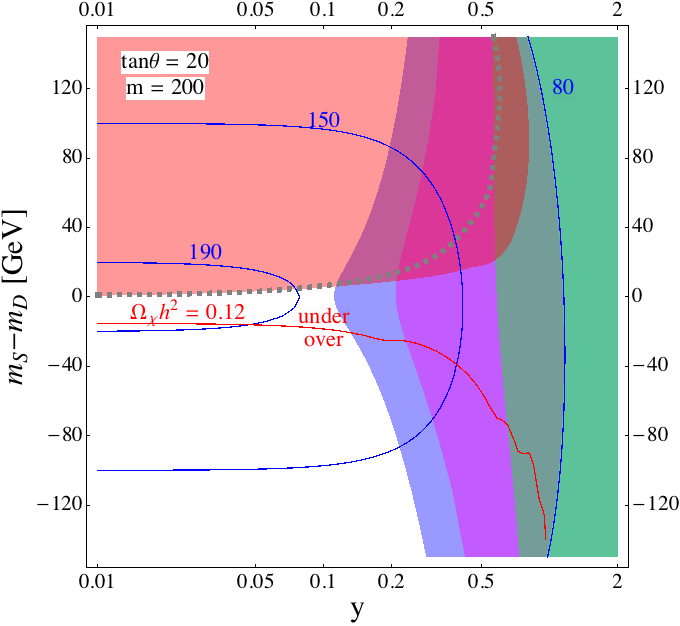}
\includegraphics[width=7cm]{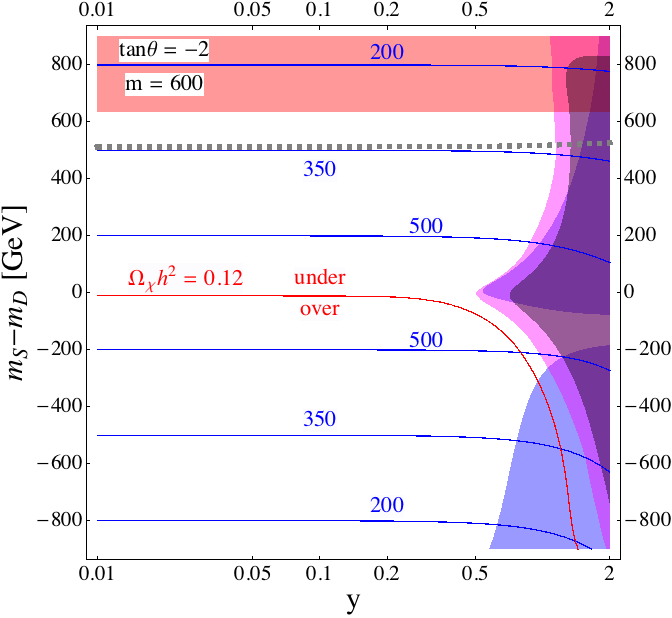}
\hspace{0.5cm}
\includegraphics[width=7cm]{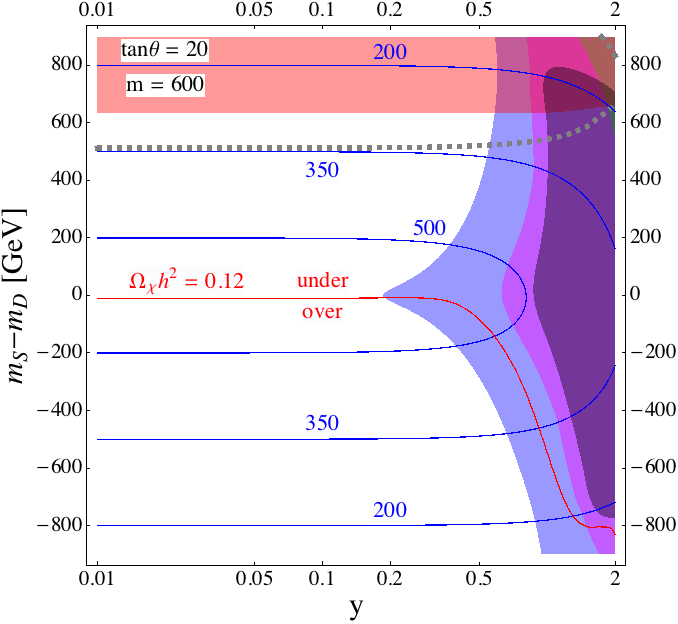}
\caption{Same as in the previous figure in the plane ($y$, $m_{\textrm{\tiny S}}-m_{\textrm{\tiny D}}$) for various choices of $\tan\theta$
and $m\equiv (m_{\textrm{\tiny S}}+ m_{\textrm{\tiny D}})/2$. The blue lines indicate contours of the DM mass $m_\chi$.}
\label{SDfer_m}
\end{figure}

The complementarity between different searches is illustrated in the aggregated plots of figures \ref{SDfer_y001}-\ref{SDfer_m}. In these plots we show the combination of bounds from indirect and direct searches as well as the prospects for the CTA experiment for a wide range of mass parameters, while the Yukawa couplings $y_{1,2}$ span from small values (we show a representative case with $y=10^{-2}$ but the constraints are the same for smaller values) and `MSSM-like' values ($y=0.2$) up to large values substantially enhancing doublet-singlet mixing effects (we choose $y=1$, a value still compatible with perturbativity up to the unification scale \cite{Mahbubani:2005pt}.). As it is clear from the mass matrix of the model, the relative hierarchy of the Yukawa couplings does not have a physical significance so it is sufficient to restrict to $|\tan\theta|\geqslant 1$.

Regarding constraints from gamma rays, as we discussed above, among all the annihilation channels the strongest bounds come from the upper limits on the annihilation cross sections $\chi\chi\to WW,ZZ$, $t\ov{t}$ and $Zh$. In the following we discuss each channel separately.

The regions excluded by Fermi-LAT for the $WW$ channel are depicted in pink in figs.~\ref{SDfer_y001}-\ref{SDfer_m}.
As discussed in the previous section, the annihilations to $WW$ and $ZZ$ are driven by electroweak interactions, hence they only depend on the doublet component of $\chi$, see eq.~(\ref{Lchi1}).\footnote{The photon fluxes from $\chi\chi\to ZZ$ are similar to those from $\chi\chi\to WW$ therefore the corresponding limits are similar. Since in our model the annihilation cross section to $WW$ is over almost all the parameter space larger than the one to $ZZ$, we show limits from $\chi\chi\to WW$.} Thus the resulting bound does not diminish in the limit of small Yukawa couplings, where it gives the only constraint to the model as shown in fig.~\ref{SDfer_y001} for $y=0.01$. In general the limits from $\chi\chi\to WW$ constrain a region of the parameter space where a thermal relic would be underabundant. Thus $\chi$ would be a viable DM candidate only under the assumption of an additional non-thermal production. If on the contrary we assume that $\chi$ is a thermal relic and therefore a subdominant DM component, one should take into account the reduced $\chi$ density and rescale the bounds by a factor $(\Omega_\chi h^2 / 0.12)^2$, resulting in practically no constraints in almost all the underabundant part of the parameter space.

In the summary plots of figs. \ref{SDfer_y001}-\ref{SDfer_y1} we have also shown the prospects for $\chi\chi\to WW$ from the CTA experiment, as estimated in \cite{Silverwood:2014yza}. It is interesting to note that the improvement is only marginal and in some regions of the parameter space (e.g.~for
$m_\chi\lesssim 200\,$GeV) the sensitivity seems to be comparable or even worse than that of Fermi-LAT. This happens because the improvement by CTA is substantial mainly in the high DM mass region where the annihilation cross section to WW is anyway too small (unless it sits on a non-perturbative resonant peak).%

The annihilation to $t\ov{t}$ is also important once this channel becomes kinematically accessible. The latest Fermi-LAT results \cite{Ackermann:2015zua} that we use do not report on limits on dark matter annihilating to $t\ov{t}$. However, since the photon flux from $t\ov{t}$ follows closely the one of $b\ov{b}$ \cite{Cirelli:2010xx}, for the purpose of demonstrating the bounds we find it sufficient to use the limits on the latter. The corresponding bounds are shown as purple regions  in figs.~\ref{SDfer_y1} and \ref{SDfer_m}.
The annihilation occurs via s-channel $Z$ boson mediation and thus the amplitude is proportional to the coupling $c_{Z\chi\chi}\sim |U_{13}|^2-|U_{12}|^2$. As we discussed below eq.~(\ref{blind}), the channel is suppressed when $\chi$ is either almost a pure doublet or pure singlet. Indeed, we see in figs.~\ref{SDfer_y001} and \ref{SDfer_y02} that for small values of $y_{1,2}$ where the mixing between the singlet and doublet neutral states is small, 
the annihilation to $t\ov{t}$, even when kinematically allowed, is highly suppressed so that it does not give any bound. 
For larger $y$, as long as $|\tan\theta|$ is not too close to 1, $\chi\chi\to t\ov{t}$ sets substantial bounds on the parameter space 
(see purple regions in fig.~\ref{SDfer_y1}), again if $\chi$ is a non-thermally produced DM candidate, as previously discussed for the $WW$ channel. These bounds, along with ones coming from the SD searches, can probe the blind-spot regions of SI searches that are controlled by the coupling to the Higgs $c_{h\chi\chi}$.

The annihilation channel $\chi\chi\to Zh$  has a relatively different behaviour than the $WW,\,ZZ$ and $t\ov{t}$ channels in the sense that its size is bounded from above by the Yukawas instead of the electroweak coupling. Therefore, it plays an important role for large values of $y$. For such $y$ the bound on SI scattering cross section from LUX becomes very strong. Nevertheless, the constrain from $\chi\chi\to Zh$ is seen to close the blind spot regions of the LUX bound. In particular, for $y=2$ and $\tan\theta=-2$ (not shown in the plots) it excludes the region $20\lesssim m_{\textrm{\tiny S}}/$GeV$\,\lesssim 400,\,200\lesssim m_{\textrm{\tiny D}}/$GeV$\,\lesssim 440$, part of which is not constrained by LUX.

The latest results of \cite{Ackermann:2015zua} include strong bounds on DM annihilating to light fermions: For $\chi\chi\to \tau^+\tau^-, b\ov{b}$, the latest limits reach as low as $\simeq 5 \times 10^{-27}~\mathrm{cm}^3\,{\mathrm s}^{-1}$ for $m_\chi\lesssim 10\,$GeV, much below the thermal relic cross section. However, due to the helicity suppression of dark matter annihilation to light fermion in our model, the cross section is generally smaller: 
$\langle\sigma v\rangle_{\tau\tau} \lesssim 3.5 \times 10^{-28}~\mathrm{cm}^3\,{\mathrm s}^{-1}$, 
$\langle\sigma v\rangle_{b\ov{b}} \lesssim 5 \times 10^{-27}~\mathrm{cm}^3\,{\mathrm s}^{-1}$. 
For this reason there are currently no bounds from these channels.

The annihilation to two photons or a photon and a $Z$ boson is maximised for $\chi$ that is mostly doublet. Similar to the supersymmetric case of a pure Higgsino DM, the thermally averaged annihilation cross section reaches a constant maximal value of 
$\langle\sigma v\rangle_{\gamma\gamma,Z\gamma}\sim 10^{-28}~\mathrm{cm}^3\,{\mathrm s}^{-1}$ in the limit of large $m_\chi$ while the full non perturbative calculation includes resonances that can increase this value by several orders of magnitude for a small mass splitting between $\chi$ and the charged state $\psi^+$ \cite{Hisano:2004ds,Bern:1997ng,Bergstrom:1997fh,Ullio:1997ke,Boudjema:2005hb}. However, in the parameter space on which we focus these effects are seen to be small enough to evade constraints on the model.

In addition to annihilation of $\chi$, the limits on spin-independent and spin-dependent scattering cross sections set strong constraints on the model, as seen in the plots of figs.~\ref{SDfer_y02}-\ref{SDfer_m}.  The SI interaction is mediated at tree level by a Higgs boson and starts to give important constraints already for moderate `MSSM-like' values of $y$, as shown in fig.~\ref{SDfer_y02}. In particular, we notice that, in contrast to indirect searches, bounds from direct detection are relevant also for thermal $\chi$ with
$\Omega_\chi h^2 \approx 0.12$, i.e.~in the region with $m_{\textrm{\tiny S}}\approx m_{\textrm{\tiny D}}$ where singlet-doublet mixing is enhanced \cite{Cheung:2013dua}. In this case LUX can set limits on $m_\chi$ up to $\sim 1$ TeV.
 Furthermore, as seen in fig.~\ref{SDfer_y1}, for large $y$ the bound from LUX becomes very strong, excluding $\chi$ as heavy as several TeV in the region of enhanced interaction, where $U_{11}\sim U_{12}\sim U_{13}$. We also note that this interaction exhibits ``blind spots", i.e. regions of the parameter space where $c_{h\chi\chi}$ becomes identically zero, as we discussed in section \ref{sec:model}. In such regions, as well as when $y$ is small ($y\lesssim 0.1$) the spin independent searches pose no bound on the model. On the other hand, the SD interaction, even  if it is relatively 
 less constrained by observations, it does not exhibit the same blind spots, as it is mediated at tree level by a $Z$ boson. Hence, similarly to $\chi\chi\to t\ov{t}$, $\sigma_{p\chi}^{SD}$ is enhanced at large $y$ as the mixing becomes bigger and broader in the parameter space. 
The bound set by PICO, shown in green, is especially relevant in the low mass region. Furthermore, as we mentioned above, for DM heavier than $m_W$ and primarily annihilating to final states that decay to hard neutrino spectra, the indirect bounds on spin-dependent cross section from IceCube can be employed and, as seen in fig.~\ref{SDfer_y1}, they lead to strong bounds depicted as magenta regions. As previously mentioned, both these bounds on $\sigma_{p\chi}^{SD}$ as well as the indirect one on $\langle\sigma v\rangle_{t\ov{t}}$ can cover certain blind spots of LUX.  This can be seen for example for $\tan\theta=-2$ in fig.~\ref{SDfer_y1}.

The coverage of the low-$y$ and blind-spot regions is also illustrated in fig.~\ref{SDfer_m}, where we see how these bounds change for different scales of $y$. The plots are made in the plane ($y$, $m_{\textrm{\tiny S}}-m_{\textrm{\tiny D}}$) for different values of $\tan\theta$ and the average singlet and doublet mass term $m\equiv (m_{\textrm{\tiny S}}+ m_{\textrm{\tiny D}})/2$.
In the top two plots we focus on light DM, taking $m=200$ GeV, corresponding to $m_\chi$ in the range 100$\div$200 GeV. 
We observe that as $y$ increases, the LUX blind spot becomes broader and for $y\gtrsim 2$, SI searches set no bound on most of the parameter space, in particular when $\tan\theta$ is negative and $m_{\textrm{\tiny S}}<m_{\textrm{\tiny D}}$. As we see, this part is covered by SD direct and indirect searches from $WW/ZZ$, $\ov{t}t$ and IceCube. The coverage of blind spots by these searches is also seen in the bottom two plots, where DM can be as heavy as $\simeq 600\,$GeV.

Putting all the complementary exclusions together allows one to set bounds on $\chi$ for a wide range of the parameter space. 
Here we focus on $m_\chi\gtrsim 100\,$GeV as we will treat the special features and details of the low mass region in a separate section below, including also constraints from colliders.
The bounds for the benchmark points of $y$ and $\tan\theta$ can be directly extracted from the contours in the plots of figs. \ref{SDfer_y001}-\ref{SDfer_y1}. Scanning over different values of $\tan\theta$ for $y\lesssim 0.01$ has no measurable effect so that one can exclude $\chi$ with $m_\chi\lesssim280\,$GeV and $U_{11}^2\lesssim0.5$ for any value of $\tan\theta$. 
For the `MSSM-like' value of $y=0.2$, the limit depends on the interplay between the SI and SD bounds as we vary $\tan\theta$, which in turn is determined by the dependence of $c_{h\chi\chi}$ and $c_{Z\chi\chi}$ on $\tan\theta$, as can be inferred from eqs.~(\ref{eq:dm_couplings1}) and (\ref{eq:dm_couplings2}). For positive $\tan\theta$ the SI bounds are strong as it can be seen in the right plots of fig. \ref{SDfer_y02}. For $\tan\theta\lesssim-1.5$ and as we decrease the value,\footnote{At the special point $\tan\theta=-1$ the mixing of the dark matter candidate practically vanishes so the situation is similar to that of the small Yukawa coupling.}
at first the dominant bound is the SD (from IceCube) but at around $\tan\theta=-4$ it gives its place to the SI bound. All in all, for $y=0.2$ we obtain the exclusion $m_\chi\lesssim220\,$GeV for $U_{11}^2\lesssim0.65$. For larger Yukawa couplings, this interplay between the various searches is even more apparent. In general we obtain stronger bounds: For $y=1$ we obtain that $\chi$ is excluded for $m_\chi\lesssim275\,$GeV unless it is almost a pure singlet ($U_{11}^2\lesssim0.8$). For positive $\tan\theta$ this bound becomes much stronger: For example, $\chi$ is excluded for $m_\chi\lesssim1\,$TeV unless it is a purely singlet or purely doublet state ($0.015\lesssim U_{11}^2\lesssim0.95$).

\subsubsection*{Exclusion on $\chi-\psi^{\pm}$ mass degeneracy}
\begin{figure}[t!]
\begin{center}
\floatbox[{\capbeside\thisfloatsetup{capbesideposition={right,top},capbesidewidth=5cm}}]{figure}[\FBwidth]
{\caption{In this illustrative scatter plot of $\sigma v_{WW}$ versus $m_\chi$ for $y=0.01$, we clearly see how the annihilation cross section to $WW$ is enhanced as the mass splitting becomes smaller. The blue line depicts the Fermi-LAT limit \cite{Ackermann:2015zua}.}}
{\includegraphics[width=9.3cm]{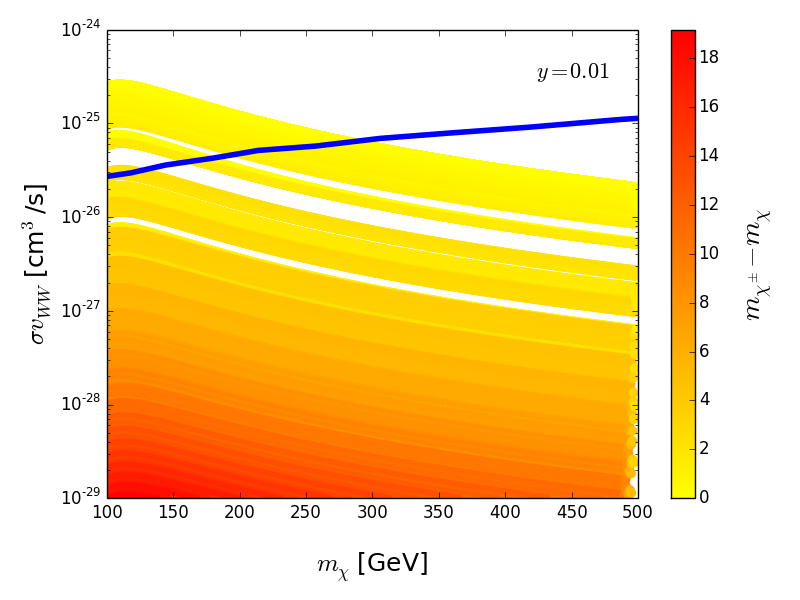}
\label{degen}}
\end{center}
\end{figure}
The more $\chi_1$ is doublet-like, the more it will have a mass that is nearly degenerate with $\chi_2$ and $\psi^+$. Therefore, the constraints on the mixing angles that we discussed above can be directly translated into exclusions of the $\chi-\psi^{\pm}$ mass splitting.

This is another set of constraints where indirect searches play a major role. Indeed, indirect searches can be used to close in on the parameter space of scenaria with compressed spectra, where other experiments such as colliders and direct searches lose sensitivity. In particular, the SI and SD cross sections are smaller when $\chi_1$ is almost degenerate with the heavier partners because the mass degeneracy is generally an outcome of $\chi_1$ being mostly doublet, so that $|U_{11}|^2\simeq 0$ and $|U_{12}|^2\simeq |U_{13}|^2\simeq 0.5$. In this case, both the $c_{\chi\chi h}\sim y_1U_{12}^*U_{11}^*+y_2U_{13}^*U_{11}^*$ and the $c_{\chi\chi Z}\sim |U_{13}|^2-|U_{12}|^2$ couplings are small and therefore direct searches cannot probe the highly degenerate regime. On the other hand, the annihilation to $WW/ZZ$ is enhanced in this regime since these channels are mediated by t-channel $\psi^\pm/\chi^0_{n}$ exchange. This behaviour is illustrated in the scatter plot of fig.~\ref{degen} where we can see that the $\sigma v_{WW}$ is maximised for nearly degenerate spectra.

Translating the bounds obtained in the previous subsection into bounds on $\delta m=m_{\psi^+}-m_{\chi}$, we have: For $y\lesssim 0.01$, $m_\chi\lesssim280\,$GeV and $|U_{11}|^2\lesssim0.5$ the exclusion is $\delta m\lesssim0.8\,$GeV. For $y=0.2$ the constrained region $m_\chi\lesssim220\,$GeV, $|U_{11}|^2\lesssim0.65$ implies the exclusion of $\delta m\lesssim20\,$GeV. For $y=1$, since the exclusion of $m_\chi\lesssim275\,$GeV holds even for $\chi$ being almost a pure singlet, the whole compressed regime is excluded. On the other hand, for positive $\tan\theta$ and $m_\chi\lesssim1\,$TeV with $0.015\lesssim|U_{11}|^2\lesssim0.95$, the lower bound on the singlet component of $\chi$ translates into exclusion of $\delta m\gtrsim15\,$GeV.

\section{Light DM regime and LHC phenomenology}
\label{sec:lhc}

In the previous sections we described the properties of the SDM and the constraints derived from DM direct and indirect detection experiments that, as we saw, can probe a very wide range of dark matter masses. In this section we would like to focus on the low DM mass region, as this is the relevant scenario for what concerns possible collider signatures. In the low mass region, besides the limits set by the DM searches, we implement other constraints on the model arising from collider experiments, in particular LEP and LHC. This makes manifest that the constraints on the model from DM searches and collider experiments are complementary, in particular in this regime.

\begin{figure}[t]
\includegraphics[width=7cm]{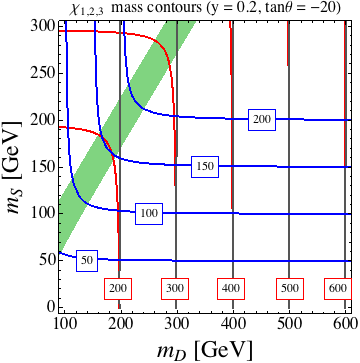}
\hspace{0.5cm}
\includegraphics[width=7cm]{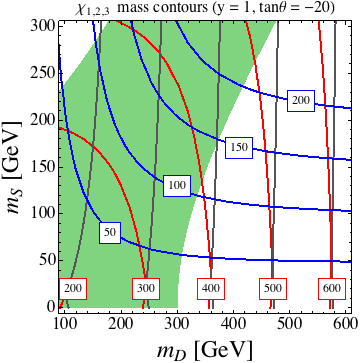}
\caption{\label{fig:mass_contours}
Mass contours for $\chi_{1}$ (blue), $\chi_2$ (gray) and $\chi_3$ (red) in the $(m_{\textrm{\tiny D}},~m_{\textrm{\tiny S}})$ plane
for different choices of $y$ and $\tan\theta$.
In the green region all three $\chi_i$ have a sizeable doublet component, see the text for details.}
\end{figure}

In order to show the behaviour of the spectrum in the low-mass region of the parameter space, 
in fig.~\ref{fig:mass_contours} we plot the mass
contours for $\chi_{1,2,3}$ in the $(m_{\textrm{\tiny D}},~m_{\textrm{\tiny S}})$ plane. 
We do not display the region with $m_{\textrm{\tiny D}}<100$ GeV, because it is excluded
by searches for charginos at LEP \cite{Abdallah:2003xe}.
We also highlight in green the regions where all three states have a sizeable doublet component and thus 
can be relevant for the LHC production modes: $\sqrt{|U_{i2}|^2+|U_{i3}|^2}>0.5$, $i=1,3$. 

For small values of $y$ (left panel of fig.~\ref{fig:mass_contours}), the second heavier neutral fermion is
always mainly doublet, with a mass close to $m_{\textrm{\tiny D}}$, such as the charged states. 
The DM candidate $\chi_1$
is mainly singlet in the region with $m_{\textrm{\tiny D}} \gg m_{\textrm{\tiny S}}$ so that $m_{\chi_1}\approx m_{\textrm{\tiny S}}$, 
while it is mainly doublet with $m_{\chi_1}\approx m_{\textrm{\tiny D}}$ for $m_{\textrm{\tiny S}} \gg m_{\textrm{\tiny D}}$. 
The $\chi_3$ mass and mixing have a complementary behaviour.
In the region $m_{\textrm{\tiny D}} \sim m_{\textrm{\tiny S}}$ the three states have comparable mass, with $\chi_2$ mainly doublet and $\chi_1$ and $\chi_3$ largely mixed. These features do not have a sensitive dependence on $\tan\theta$.
Such simple pattern is highly distorted for large values $y$, as shown in the right panel of fig.~\ref{fig:mass_contours},
as  a consequence of the large off-diagonal entries in the mass matrix. In particular, large mixing as well as 
large mass splittings among the three neutral states arise (and with respect to the charged ones). This might lead to interesting
signatures at colliders, as we will discuss below.

\subsection{Dark Matter and EW constraints}
In this subsection we focus on possible modifications induced by the new physics sector to the measured properties of the SM particles.
In particular, we quantify the contributions to EW precision observables (EWPO) and the invisible decay widths of the $Z$ and $h$.
\begin{itemize} 
\item For what concerns EWPO, we use the expressions in  \cite{Barbieri:2006bg,Eramo:2007ab}
to compute the new physics contributions to the oblique parameters, $\Delta S$ and $\Delta T$. 
As already observed in \cite{Enberg:2007rp}, the strongest constraints to the model arise from $T$, while
the contribution to $S$ is always small.
Since $\Delta T$ scales as $\sim (y_1^2-y_2^2)^2$, EWPO constraints can be relevant in scenarios with large $y$ and large $\tan \theta$.
Given the small contribution to $S$, we simply impose the bound $|\Delta T|<0.2$  \cite{Baak:2014ora}.
\item In the SDM, the $Z$ boson width can receive relevant corrections if the decay $Z \to \chi_1 \chi_1$ in kinematically accessible. 
The contribution to the invisible width of the Z boson is
\be
\Gamma(Z \to \chi_1 \chi_1)= \frac{m_Z}{6\pi} \left(1 -\frac{4 m_{\chi_1}^2}{m_Z^2}\right)^{\frac{3}{2}} \left|c_{Z\chi\chi}\right|^2\,,
\ee
where the $Z\chi_1\chi_1$ coupling is given in eq.~(\ref{eq:dm_couplings2}).
The $Z$ width measurement performed by LEP  
impose the following constraint on possible new physics contribution to $\Gamma(Z \to \text{invisible})$: 
$\Delta\Gamma_Z^{\text{inv}} < 3 \text{ MeV}$  (95\%~{CL}) \cite{ALEPH:2005ab}.
\item Analogously, the Higgs boson width into invisible particles can also receive important corrections in the SDM
if $m_{\chi_1}<m_h/2$. The Higgs boson partial width to $\chi_1$ reads
\be
\Gamma(h \to \chi_1 \chi_1)= \frac{m_h}{4 \pi} \left(1 -\frac{4 m_{\chi_1}^2}{m_h^2}\right)^{\frac{3}{2}}
\left| c_{h\chi\chi}\right|^2\,,
\ee
where the coupling is as in eq.~(\ref{eq:dm_couplings1}).
New invisible decay modes of the Higgs would reduce all visible branching fractions, hence the invisible 
branching ratio can be constrained by fits to the observed decay rates.
In the following we employ the following limit: 
${\text{BR}}(h \to \text{invisible})  < 26\%$ (95\%~{CL})  \cite{Bechtle:2014ewa}.
\end{itemize}
\begin{figure}[t]
\includegraphics[width=7cm]{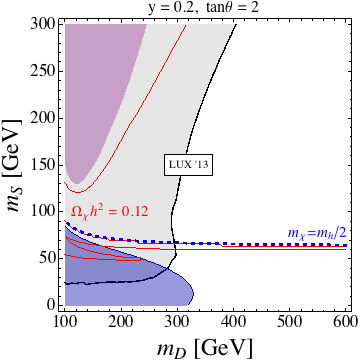}
\includegraphics[width=7cm]{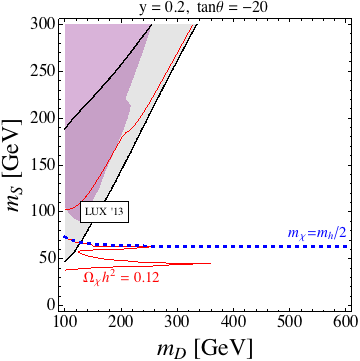}
\caption{\label{lowmass_bound_1}
Constraints on SDM in the low-mass range for $y=0.2$ and $\tan\theta=2,\,-20$. 
Gray: region excluded by direct detection experiments. 
Purple: region excluded by indirect detection experiments. 
Blue: exclusion from ${\text{BR}}(h \to \text{invisible})$.}
\end{figure}

\begin{figure}[t]
\includegraphics[width=8cm]{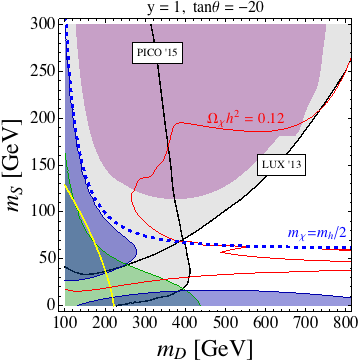}
\caption{\label{lowmass_bound_2}
Same as in figure \ref{lowmass_bound_1} for $y=1$ and $\tan\theta=-20$. 
Additional excluded regions that appear are the following.
Green: exclusion from $\Delta\Gamma_Z^{\text{inv}}$.
Yellow: limit from EWPO.}
\end{figure}
In the figures \ref{lowmass_bound_1} and \ref{lowmass_bound_2} we show the above constraints together with those
from DM experiments discussed in the previous sections for the low-mass region of the parameter space.
Regions excluded by direct detection are shown in gray, while the indirect detection bounds are collectively marked in purple.
As before, the red line denotes $\Omega_{\chi}h^2= 0.12$.  
From the figures, one can observe the resonant enhancement of the DM annihilation cross section when
$m_{\chi_1}\approx m_Z/2$ and $m_{\chi_1}\approx m_h/2$, respectively due to $Z$ and $h$ s-channel exchange.
For these configurations, an efficient DM annihilation can be obtained even for a mostly singlet-like $\chi_1$, 
i.e.~$m_{\textrm{\tiny D}}\gg m_{\textrm{\tiny S}}$. 

In figure \ref{lowmass_bound_1}, we show the case $y=0.2$ that resembles the Higgsino-Bino system of the MSSM, 
as discussed in the previous sections. 
Given the moderate value of the couplings, the EWPO do not pose any bound, whereas the Higgs invisible width constraint (blue-shaded area)
can exclude regions with small $m_{\chi_1}$ in the case of low $\tan \theta$. 
Larger and/or negative values of $\tan \theta$ can reduce or even evade this bound, as shown in the right panel of the figure.
This is due to the dependence of the  $h \chi_1\chi_1$ coupling on $\tan \theta$ that we can obtain from eq.~(\ref{eq:dm_couplings1}).
Furthermore, we see that there is a  partial cancellation between the two terms of $c_{h \chi\chi}$ in case $y_1$ and $y_2$ have opposite signs.
This is also responsible for the reduction of the direct detection bound in the right panel of the figure, as well as the decreased efficiency
of DM annihilation on the $h$ resonance \cite{Calibbi:2014lga}.

The case of large $y$ is strongly constrained by direct searches for dark matter. 
In particular, for moderate and/or positive values of $\tan\theta$, the LUX bound can be evaded only for $m_{\text{\tiny D}}$ in the multi-TeV range.
For this reason, we only show the case $y=1$, $\tan\theta=-20$, see figure \ref{lowmass_bound_2}.
As we can see, limits on invisible $Z$ and $h$ decays set relevant constraints for low values of $m_{\textrm{\tiny S}}$
and $m_{\textrm{\tiny D}}$, depicted respectively as green and blue regions. The bound from $\Delta T$ (shown as a yellow line) is also relevant although somewhat weaker.
As mentioned above, DM experiments are very constraining, especially because of the increase of the couplings $c_{h\chi\chi}$ and  $c_{Z\chi\chi}$ at large $y$.
However, we stress that, as discussed in the previous section, the bounds obtained from DM experiments 
assume that 100\% of the observed dark matter is accounted for by our candidate $\chi_1$. Thus 
they can be relaxed in the regions where a thermal $\chi_1$ would give $\Omega_\chi h^2 \ll 0.12$, assuming that it is only a subdominant component of the total DM abundance. In particular, in the special case $m_{\chi_1} \simeq {m_h}/{2}$, where dark matter annihilation process is resonantly enhanced, the bounds from direct and indirect searches can be evaded to large extent. 
For this reason, in the following we also discuss a distinctive collider signature that can be realised in such peculiar region of the parameter space (together with the more standard LHC phenomenology of the moderate $y$ case).
 
As a final remark, we can now complete the discussion of subsection \ref{results} setting comprehensive bounds also in the low mass region, considering at the same time all the experimental constraints: Combining both analyses, we see that for $y=0.2$, one can exclude $\chi$ for $80\,$GeV\,$\lesssim m_\chi\lesssim220\,$GeV for $U_{11}^2\lesssim0.65$. For the large Yukawa benchmark ($y=1$), the bounds from invisible decays allow to keep the bound of the previous section ($m_\chi\lesssim275\,$GeV for $U_{11}^2\lesssim0.8$) even for $\chi$ of a few GeV.

\subsection{LHC signatures and constraints}
In this subsection we discuss the LHC phenomenology of the model.
A detailed analysis and a recasting of existing LHC searches to this scenario are beyond the scope of this work. Here we restrict to a discussion of the generic features that are characteristic of different parameter space regions. An early study of the LHC signatures and prospects of the model
can be found in \cite{Enberg:2007rp}.
As discussed in section \ref{sec:model}, the SDM can be considered as a generalisation of the Higgsino-Bino system of the MSSM.
This analogy is very useful in exploring the collider phenomenology of the model, 
since the supersymmetric case has been extensively discussed in the literature  \cite{Bharucha:2013epa,Gori:2013ala,Han:2013kza,Schwaller:2013baa,Han:2014kaa,Han:2014xoa,Martin:2014qra,Calibbi:2014lga,Bramante:2014tba,diCortona:2014yua}, with the important difference that here the mass splittings are controlled by $y$ instead of $g'$ and thus can be substantially larger than in the MSSM.

The main production mechanisms at LHC are the electroweak Drell-Yan processes, 
i.e.~charged fermion pair production through s-channel $Z$-boson or photon exchange,
associated production of charged and neutral fermions through s-channel $W$
and neutral fermions pair production through $Z$:
\be
p p \to \psi^+ \psi^- ~,\qquad 
p p \to \chi_i \psi^{\pm}  ~ ,\qquad p p \to \chi_i \chi_j ~, \qquad i,j= 1,2,3\,. 
\ee
The production involving neutral states is controlled by their couplings to the $W$ and $Z$ bosons that, as we have seen, 
are weighted by their doublet components, i.e.~by entries of the mixing matrix $U$, see eq.~(\ref{eq:couplings}). 
In principle each of the three neutral fermions can have a large doublet component, independently of the mass hierarchy among them, as has
been highlighted in figure \ref{fig:mass_contours}.
The figure encodes the information about the dominant production modes we expect at LHC through the parameter space.
Within the green bands all three neutral states can be substantially produced (obviously if their masses do not exceed the few-hundreds GeV range), while on the right (left) of green regions the main production modes involve the two heavier (lighter) states 
$\chi_2,\chi_3$ ($\chi_1,\chi_2$).
\begin{figure}[t!]
\includegraphics[width=7cm]{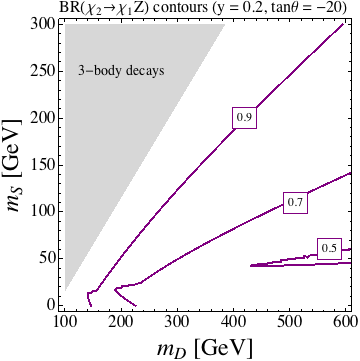}
\hspace{0.5cm}
\includegraphics[width=7cm]{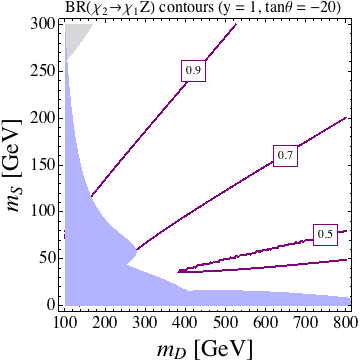}
\includegraphics[width=7cm]{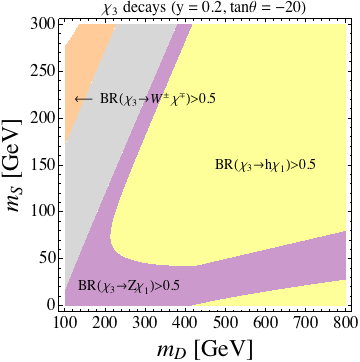}
\hspace{0.5cm}
\includegraphics[width=7cm]{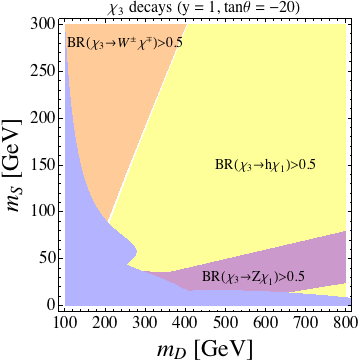}
\caption{\label{BR_plots}
Heavy neutral states decays for $y=0.2$ (left) and $y=1$ (left) $\tan\theta=-20$.
First row: contours of BR$(\chi_2 \to \chi_1 Z)$. Second row: regions showing the dominant decays of $\chi_3$.
}
\end{figure}
The final state topology depends on the decay modes of the produced particles.
The charged fermions will predominantly decay to $\chi_1$ and a (possibly off-shell) $W$-boson.
This decay occurs trough the (even tiny) doublet component of $\chi_1$ and it will be prompt in the mass 
range we consider, as soon as
$U_{12}$ or $U_{13} \gtrsim \mathcal{O}(10^{-8})$, i.e.~$y\gtrsim  \mathcal{O}(10^{-7})$.
The decay modes of the mostly-doublet neutral fermions can instead present a rich pattern.
The heaviest can have the following decay modes:
\be
\label{decays_n3}
\chi_3 \to W^{\pm} \psi^{\mp} \quad, \quad 
\chi_3 \to h \chi_2~, \quad \chi_3 \to \chi_2 Z~, \quad \chi_3 \to  \chi_1 h~, \quad \chi_3 \to \chi_1 Z
\ee
Among these, the decay into $\psi^+$ and $W$ is especially interesting. 
When this decay is kinematically open, the large coupling to $W$, c.f.~eq.~(\ref{eq:couplings}), 
makes it the dominant mode leading to characteristic  collider signatures, as we discuss below.
The second heavier neutral fermion mainly decays to the lightest neutral fermion: 
\be
\chi_2 \to \chi_1 Z~, \quad \chi_2 \to \chi_1 h
\ee
and possibly to the charged fermion in very peculiar corners of the parameter space, where $m_{\chi_2} > m_W + m_{\textrm{\tiny D}}$.
Analytical formulae for the tree-level decay widths are reported in Appendix \ref{app:decay}.
Obviously if the above modes are kinematically forbidden, the decays will occur through off-shell $W$, $Z$ or $h$,
depending on the coupling strength.

In Figure \ref{BR_plots} we show the dependence of the branching ratios
on $m_{\textrm{\tiny D}}$ and $m_{\textrm{\tiny S}}$ for $y=0.2$ (left) and $y=1$ (right) and $\tan\theta=-20$.
In the first row we plot contours of BR$(\chi_2 \to \chi_1 Z)$, in the second row we show regions corresponding to 
the dominant decay modes of $\chi_3$. Gray-shaded areas highlight regions where only three-body decays are kinematically accessible, such
that a reduction of the LHC sensitivity is expected. The blue-shaded regions are excluded by the collider constraints discussed in the 
previous subsection, namely EWPO, $Z$ and Higgs boson invisible decay widths.\\

The most interesting and easiest to probe final states at colliders are the ones involving the production of gauge bosons, that we will discuss
in more detail in the following. Indeed, they can lead to distinctive and clean signatures at LHC, such as multi-leptons plus missing energy.
We will focus here on the leptonic decays of the gauge bosons and classify the possible signatures in terms of number, sign and flavours 
of leptons in the final state.\footnote{We consider only the cases corresponding to dominant combinations of productions and branching fractions. 
For instance, we neglect modes involving $\chi_3 \to \chi_2 Z$, as $\chi_3 \to \psi^\pm W^\mp$ will typically dominate in the regions where both
decays are kinematically open. Furthermore, for doublet-like $\chi_2$ and $\chi_3$, 
the combination of mixing angles in $c_{Z\chi_m\chi_n}$, cf.~eq.~(\ref{zetaeta}),
makes the production of $\chi_3 \chi_2$ always larger than $\chi_3 \chi_3$ and $\chi_2 \chi_2$, hence we only consider 
$pp\to \chi_3 \chi_2$ among the neutral fermion production modes.} 

\paragraph{2 leptons.}
Events with two (possibly different-flavour) opposite-sign leptons plus missing energy
can follow the charged fermion pair production
\be
pp\to \psi^+ \psi^- \to W^+  \chi_1 W^- \chi_1\,,
\ee
exactly like in the case of chargino production in the MSSM.
Due to the large SM background, this is one of the few electroweak production 
modes for which LHC experiments have not
set stronger constraints than LEP yet.
\paragraph{3 leptons, with a pair reconstructing the $Z$-boson mass.}
This final state leads to the most stringent LHC constraints at the moment.
The production modes contributing to this signature are 
\be
p p \to \chi_{2,3}\psi^{\pm} \to Z \chi_1 W^{\pm} \chi_1\,.
\ee
Hence schematically, the $W$ plus $Z$ plus missing energy production rate will be given by
\be
\sigma (\chi_3 \psi^\pm)\times {\rm BR}(\chi_3 \to Z \chi_1) + \sigma (\chi_2 \psi^\pm) \times {\rm BR}(\chi_2 \to Z \chi_1) \,. 
\ee
Searches based on this kind of events have been published in \cite{Aad:2014nua,Khachatryan:2014qwa}. 
The ATLAS search \cite{Aad:2014nua} has been recently recasted in the context 
of the Higgsino-Bino system of the MSSM \cite{Calibbi:2014lga}. 
We can use this reinterpretation to estimate the LHC sensitivity to our parameter space.
From the analysis of \cite{Calibbi:2014lga}, we see that the ATLAS search \cite{Aad:2014nua} 
is sensitive to our model roughly under the following conditions: 
when $m_{\chi_{2}},m_{\chi_{3}},m_{\psi^+}\lesssim$ 270 GeV, $m_{\chi_1}\lesssim 75$ GeV
and the average of ${\rm BR}(\chi_2 \to Z \chi_1)$ and ${\rm BR}(\chi_3 \to Z \chi_1)$ is at least $60\%$.
Let's consider for instance the case $y=0.2$, $\tan\theta=-20$ .
From the mass contours and the branching ratios shown in figures~\ref{fig:mass_contours} and \ref{BR_plots}, 
we see that the region relevant for the $Z$ and $h$ resonances in figure \ref{lowmass_bound_1}
with $m_{\textrm{\tiny D}} \lesssim 270$ GeV is likely excluded by the ATLAS search \cite{Aad:2014nua}.
For the other cases shown in figure \ref{lowmass_bound_1} and \ref{lowmass_bound_2}, the bound coming
from BR($h\to \chi_1\chi_1$) seems to be at present more stringent than the 3-lepton search.
\paragraph{3 leptons, without a pair reconstructing the $Z$-boson mass.}
This final state can be obtained from decays of three $W$ bosons. 
The production mechanism is $\chi_3 \psi^{\pm}$ with $\chi_3$ decaying to $W^\pm \psi^\mp$:
\be
p p \to \chi_3  \psi^{\pm} \to W^\pm \psi^\mp W^{\pm} \chi_1 \to  W^{\pm} W^{\mp} \chi_1  W^{\pm} \chi_1 \,.  
\ee
This case can be realised if $m_{\chi_3} > m_W +m_{\textrm{\tiny D}} $ and $\chi_3$ has a sizeable doublet component.
These conditions are difficult to obtain for low $y$, as we can see comparing the plots of figures~\ref{fig:mass_contours} and \ref{BR_plots}:
in the region where the decay of $\chi_3$ into $W$ is preferred, the doublet component of $\chi_3$ is small, reducing the production cross section.
Instead, this signature can be realised in the case of large $y$. For instance, consider the region of the plane in figure 
\ref{lowmass_bound_1} where $m_{\chi_1} \approx{m_h}/{2}$ in the
case of $y=1$ and $\tan \theta=-20$, e.g.~with $m_{\textrm{\tiny S}} \simeq m_{\textrm{\tiny D}} \simeq 150$ GeV. There $\chi_3$ is mostly doublet and decays preferably to $W^\pm \psi^\mp$, realising the scenario we just described.

\paragraph{4 leptons with one pair reconstructing the $Z$-boson mass.}
This final state can be obtained via the process:
\be
p p \to \chi_3 \chi_2 \to W^{\pm} \psi^{\mp} Z \chi_1  \to  W^{\pm}  W^{\mp} \chi_1 Z \chi_1
\ee
This case can also be realised in the same region of the parameter space of the $y=1$ and $\tan \theta=-20$ case discussed above, just close to the line
of Higgs enhanced annihilation in figure \ref{lowmass_bound_1}.
Indeed there $\chi_2$ is mostly doublet and decays preferably to $Z$ plus $\chi_1$, see figure \ref{BR_plots}.
The final state signature is two $W$ bosons (of opposite sign) plus one $Z$ and missing energy.
Two out of the four leptons will have same flavour and opposite charge and reconstruct the $Z$ boson mass, 
while the other two will have opposite charge but of uncorrelated flavour.

\paragraph{4 leptons with two pairs reconstructing the $Z$-boson mass.}
This case follows from the same production mode as the previous one when $\chi_3$ also undergoes the more ordinary decay to $Z$: 
\be
p p \to \chi_3 \chi_2 \to Z\chi_1 Z \chi_1 \,. 
\ee
Considering the behaviour of the branching ratios shown in figure \ref{BR_plots}, this mode is particularly relevant in the regions with 
small $m_{\textrm{\tiny S}}$ and $m_{\textrm{\tiny D}}$ in the few hundred GeV range. This makes it, together with the 3-leptons mode from $WZ$ discussed above, very promising to test the unconstrained regions in figure \ref{lowmass_bound_1} and \ref{lowmass_bound_2} at the current
LHC run.

\section{Summary}
\label{summary}

In the present work we have analysed the theory and the experimental constraints of a dark matter candidate that is a mixture of an electroweak doublet and singlet fermion. The model enjoys certain properties: i) It represents a minimal viable description of electroweak dark matter that is stable due to a symmetry of the theory. ii) The currently running direct detection experiments are sensitive enough to probe  DM-nucleon interactions via Higgs exchange. Since this model has tree level DM-Higgs interactions, it can be used as a simplified model to parametrise the bounds on this coupling. iii) It improves gauge coupling unification of the SM. iv) It captures the relevant dynamics of DM candidates of UV theories such as the Bino-Higgsino system in MSSM and the Singlino-Higgsino system in NMSSM-type scenarios.

We have used a combination of results from direct search experiments (LUX \cite{Akerib:2013tjd}, PICO \cite{Amole:2015lsj}), indirect search experiments (FERMI \cite{Ackermann:2015zua}, ICECUBE \cite{Aartsen:2012kia}) and collider searches (Higgs and $Z$ invisible decays as well as EW precision measurements) in order to set constraints on the viable parameter space of this model.

We focused on three different regimes for the DM-Higgs Yukawa coupling $y$: Small ($y=0.01$), MSSM-type ($y=0.2$) and large ($y=1$). Lowering $y$ below $0.01$ does does not change substantially the phenomenology of the model while values relatively larger than $1$ are not compatible with perturbativity up to the unification scale.

We have seen that, while colliders can probe DM candidates up to a few hundred GeV, direct and indirect searches allow us to extend the limits up to masses of several TeV, depending on the composition of DM. The bounds on the parameter space from direct and indirect DM searches are summarised in figures \ref{SDfer_y001} - \ref{SDfer_y1}, while the combined bounds from all experiments in the low mass regime are shown in figures \ref{lowmass_bound_1} and \ref{lowmass_bound_2}. For small $y$ ($\lesssim 0.1$) the FERMI limits on $\chi\chi\to WW$ exclude $\chi$ that is mostly doublet up to $\simeq 280\,$GeV, while for $y=0.2$, SI, SD and collider limits are also added and allow one to exclude the DM candidate $\chi$ with a singlet component $U_{11}^2 \lesssim 0.65$ for masses $\lesssim 230\,$GeV. For $y=1$, SI direct searches set strong bounds on the model, apart from regions of the parameter space where the DM-Higgs coupling is vanishing. We have shown that these `blind-spots' are probed by other searches, in particular SD and indirect searches. This allows to exclude $\chi$ with mass $\lesssim 275\,$GeV unless it is a highly pure singlet state ($U_{11}^2\gtrsim 0.8$). If we restrict to positive values for the Yukawa couplings, $\chi$ is excluded up to several TeV unless it is an almost pure state.

We have also discussed the phenomenology of the model at LHC. The signatures are similar to the more studied Bino-Higgsino system with the important difference that the coupling to the Higgs and the mass splittings are controlled by $y$ instead of the hypercharge coupling. The most interesting final states involve 2, 3 or 4 leptons plus missing energy. It would be interesting to perform a detailed study of such signatures at LHC at $\sqrt{s}=13\,$TeV and in particular to ascertain the reach for the region of parameter space that is still allowed in the moderate-low mass case (see the white region of figure \ref{lowmass_bound_2}).

Regarding future directions, in this work we have not included constraints coming from limits on fluxes of charged cosmic ray particles. It would be interesting to see how these limits can be used in combination with the present ones in order to further constrain the model. Also, the computation of DM-nucleon scattering amplitude has been done at tree level. A more detailed treatment would require the inclusion of loop corrections that can be relevant \cite{Hill:2014yka} in regions of the parameter space where the tree level coupling is small, such as the pure doublet limit.

\section*{Acknowledgements}

The authors would like to thank B. Zaldivar for useful discussions. PT and AM are supported in part by Vrije Universiteit Brussel through the Strategic Research Program `High-Energy Physics', in part by the Belgian Federal Science Policy Office through the Inter-university Attraction Pole P7/37 and in part by FWO-Vlaanderen through project G011410N. AM is a Pegasus FWO postdoctoral Fellow.

\newpage

\appendix

\section{Decay rates}
\label{app:decay}

In this appendix we report the analytic formula \cite{Drees:873465} for the decay of the charged fermion and the heavy neutral fermions used in the main body of the paper.

Given a generic Yukawa-type interaction of the form $ \bar \psi (c_L P_L + c_R P_R) \chi \phi$, the partial width for the decay $\chi \to \psi \phi$ is given by
\be
\Gamma (\chi \to \psi \phi) =  \frac{\lambda (m_{\chi}^2,m_{\psi}^2,m_{\phi}^2)^{1/2} }{32 \pi m_{\chi}^3} \left(
(|c_L|^2+|c_R|^2)(m_{\chi}^2+m_{\psi}^2-m_{\phi}^2)+ 2 m_{\psi}^2 (c_R c_L^*+ c_L c_R^*)
\right),
\ee
where $\lambda(a,b,c)=a^2+b^2+c^2-2 a b -2 ac - 2 bc$.

Given a generic vector boson interaction $ g V^{\mu} \bar \psi \gamma^{\mu} (c_L P_L + c_R P_R) \chi$ the partial width for the decay $\chi \to V \psi $ is given by
\bea
\Gamma (\chi \to V \psi)&=&
\frac{g^2 \lambda (m_{\chi}^2,m_{\psi}^2,m_V^2)^{1/2} }{32 \pi m_{\chi}^3} 
\left[
-6 (c_R c_L^*+ c_L c_R^*) m_{\chi} m_{\psi}+ 
\right.
\nonumber
\\
&&
\left.
(|c_L|^2+|c_R|^2) (m_{\chi}^2+m_{\psi}^2 - m_V^2 + ((m_{\chi}^2-m_{\psi}^2)^2-m_V^4)/m_V^2) 
\right].
\eea

We are interested in the possible decay of the mass eigenstate fermions in the singlet-doublet model described in Section \ref{sec:model}.
The coupling of the charged fermion with the three neutral fermions and the $W$ boson is given by 
\be
(c_L)_i= -\frac{1}{\sqrt{2}} U_{i,3}\,, \qquad (c_R)_j=\frac{1}{\sqrt{2}} U^*_{j,2}\,,
\ee
where here $i$ refers to the i-th mass eigenstate of the neutral fermions, ordered in absolute value of the mass.
The couplings among the neutral fermions and the $Z$-bosons are
\be
(c_L)_{ij} =-\frac{1}{2} (U^*_{i,2} U_{j,2}-U_{i,3}^* U_{j,3})\,,  \qquad (c_R)_{ij}=-(c_L^*)_{ij}\,. 
\ee
Finally, the coupling to the Higgs boson are
\footnote{Compared to the expression in the main body of the paper, we have symmetrized and taken into account a $1/2$ factor in the Lagrangian.}
\be
(c_L)_{ij}= -\frac{1}{\sqrt{2}}\left( y_1 U^*_{i1} U^*_{j2} +y_1 U^*_{j1} U^*_{i2} +y_2 U^*_{i1} U^*_{j3}+y_2 U^*_{j1} U^*_{i3} \right)\,, 
\qquad (c_R)_{ij}= (c_L^*)_{ij}\,.
\ee

\newpage

\bibliography{coannihilation}{}
\bibliographystyle{h-physrev}
\newpage

\end{document}